\documentclass[12pt, border=55mm,preprintnumbers]{article}

\usepackage{amssymb}
\usepackage{amsmath}
\usepackage{slashed}
\usepackage{graphicx}
\usepackage{color}
\usepackage{relsize}
\usepackage{lmodern}
\usepackage{hyperref}  
\usepackage{float}

\def\bea{\begin{eqnarray}}
\def\eea{\end{eqnarray}}
\def\bean{\begin{eqnarray*}}
\def\eean{\end{eqnarray*}}

\begin{document}

\thispagestyle{empty}

\noindent\
\\
\\
\\
\begin{center}
\large \bf Grand Unified Theory with a Stable Proton\footnote{Plenary talk given at the Conference on Particles and Cosmology, Singapore, March 5--9, 2018;
based on the work: B.~Fornal and B.~Grinstein,
Phys.\,Rev.\,Lett.~119, 241801 (2017) \cite{Fornal:2017xcj}; speaker: B.~Fornal.}
\end{center}
\hfill
 \vspace*{1cm}
\noindent
\begin{center}
{\bf Bartosz Fornal and Benjam\'{i}n~Grinstein}\\ \vspace{2mm}
{\emph{Department of Physics, University of California, San Diego \\
9500 Gilman Drive, La Jolla, CA 92093, USA}}
\vspace*{1.5cm}
\end{center}

\begin{abstract}
We demonstrate that a phenomenologically viable four-dimensional  grand unified theory with no proton decay can be constructed. This is done in the framework of the minimal non-supersymmetric ${\rm SU}(5)$ GUT by introducing new representations and separating the physical quark and lepton fields into different multiplets. In such a theory all beyond Standard Model particles are naturally heavy, but one can tune the parameters of the model such that gauge coupling unification is achieved and some of the new particles are at the TeV scale and accessible at the LHC.
\end{abstract}

\newpage

\section{Introduction}
The idea of grand unification was proposed shortly after the Standard Model (SM) of elementary particles was completely formulated based on the gauge group ${\rm SU}(3)_c\times {\rm SU}(2)_L \times {\rm U}(1)_Y$ \cite{Glashow:1961tr,Weinberg:1967tq,Salam:1968rm,SU(3),Fritzsch:1973pi}. Grand unification postulates that the three gauge interactions of the SM -- the electromagnetic, weak, and strong forces -- are the manifestation of a single force at high energies. 
The first partially unified  theory was the Pati-Salam model built on the gauge group ${\rm SU}(4)\times {\rm SU}(2)_L \times {\rm SU}(2)_R$ \cite{Pati:1974yy}. Subsequent proposals of complete grand unification were based on  ${\rm SU}(5)$   \cite{Georgi:1974sy} and ${\rm SO}(10)$ \cite{Fritzsch:1974nn,Georgi}. 

Grand unified theories (GUTs) are the holy
grail of particle physics, bringing orderliness to the otherwise unrelated particles and interactions of the SM. 
For the last 40 years it has been commonly believed that in any realistic four-dimensional (4D) GUT the
proton cannot be stable. Increasingly stringent experimental bounds on the proton lifetime \cite{Miura:2016krn} severely constrained existing GUTs, often excluding their minimal realization \cite{Nath:2006ut,Dorsner:2012nq}. Thus, many have been led to consider instead theories without a
single unifying gauge group, loosing the most appealing property of GUTs -- complete unification. 

We have shown by an explicit construction that 4D GUTs with a stable proton based on a
single gauge group that are phenomenologically viable do {in fact exist \cite{Fornal:2017xcj}. A discussion of this is presented below. \vspace{1mm}

\section{Minimal SU(5)}
Since our model is  based  on the ${\rm SU}(5)$ gauge group,  we first review briefly the key elements of the minimal ${\rm SU}(5)$ GUT -- its particle content, Lagrangian, symmetry breaking pattern and proton decay channels.

\subsection{Fermion sector}
There are two fermion irreducible ${\rm SU}(5)$ representations (irreps) containing all SM matter fields of a given family. In terms of left-handed fields these are the $5^c$ and $10$, where ``$c$'' denotes charge conjugation. The decomposition of those ${\rm SU}(5)$ multiplets into representations of the SM gauge group is (for simplicity, we consider only the first generation): 
\bean
5^c= l \oplus d^c  \ , \ \ \ \ \ 10 = e^c \oplus q \oplus u^c \ ,
\eean

\noindent
where $l$ and $q$ are the SM left-handed lepton doublet and quark doublet, respectively,  while $e$, $d$ and $u$ are the SM right-handed electron, down quark and up quark.
The explicit decomposition including the ${\rm SU}(3)_c$ and ${\rm SU}(2)_L$ indices is provided in the appendix.

\subsection{Higgs sector and symmetry breaking}
The two scalar irreps in the minimal ${\rm SU}(5)$ model are:
\bean
5_H \!\!\!&=&\!\!\! H \oplus (3,1)_{-1/3} \ ,\nonumber\\
24_H \!\!\!&=&\!\!\! (1,1)_0 \oplus (1,3)_0 \oplus (3,2)_{-5/6} \oplus (\bar{3},2)_{5/6} \oplus (8,1)_0 \ .
\eean
Assuming a $\mathcal{Z}_2$ symmetry of the Lagrangian under $24_H \rightarrow - 24_H$, the part of the scalar potential involving just the adjoint $24_H$ takes the form
\bean
V(24_H) = -\tfrac{1}{2} \mu_{24}^2 \,{\rm Tr}\!\left(24_H^2\right) + \tfrac{1}{4} a_1 \!\left[{\rm Tr}\!\left(24_H^2\right) \right]^2 + \tfrac{1}{4} a_2  \,{\rm Tr}\!\left(24_H^4\right) .
\eean
The $24_H$ develops a vacuum expectation value (vev) at the GUT scale,
\bean
  \langle \,24_H \rangle = \tfrac{1}{\sqrt{30}}\,v_{24} \,
  {\rm diag}\left(
  2, 2, 2, -3, -3
  \right),
\eean
which spontaneously breaks the symmetry ${\rm SU}(5) \rightarrow {\rm SU}(3)_c\times {\rm SU}(2)_L \times {\rm U}(1)_Y$.  The fields $(3,2)_{-5/6}$ and $(\bar{3},2)_{5/6}$ are the would-be Goldstone bosons of the broken ${\rm SU}(5)$. The other fields in the $24_H$ obtain masses on the order of $v_{24}$ and $\mu_{24}$, thus they are all at the GUT scale.

The SM Higgs doublet in the $5_H$  develops the standard electroweak vev,
which further breaks ${\rm SU}(2)_L \times {\rm U}(1)_Y \rightarrow {\rm U}(1)_{\rm em}$. For the most general form of the scalar potential $V(5_H, 24_H)$ the doublet and triplet in $5_H$ generically have masses of the order of the GUT scale, and a tuning of parameters is required for the  SM Higgs mass to be down at the electroweak scale. This is known as the doublet-triplet splitting problem.

\subsection{Gauge bosons}
In a theory based on ${\rm SU}(5)$ there are 24 gauge bosons, $A_\mu^a$, where $a=1, ..., 24$.  
Upon ${\rm SU}(5)$ breaking, those gauge bosons become the 8 gluons, 4 electroweak gauge bosons and the heavy vector gauge bosons $X_\mu = (3,2)_{-5/6}$ and  $\overline{X}_\mu = (\bar3,2)_{5/6}$ with mass
\bean
m_X = \sqrt{\tfrac{5}{6}}\, g\, v_{24} \ ,
\eean
where $g$ is the ${\rm SU}(5)$ gauge coupling constant.

\subsection{Quark and lepton masses}
The Yukawa sector of the minimal ${\rm SU}(5)$  is given by
\bean
\mathcal{L}_Y \, = \,   y_5 \ 5^c \, 10 \ 5^*_H  \,+ \,   y_{10}\,  10\ 10 \ 5_H \ \supset \    y_5 \, l\, H^*  e^c +   y_5 \, q\, H^*  d^c + y_{10} \, q\, H\,  u^c 
\eean
and results in the prediction $m_e = m_d$, $m_\mu = m_s$ and $m_\tau = m_b$ at the GUT scale. While the relation $m_\tau = m_b$, after running down to the low scale, is roughly consistent with experimental data, the relations $m_e = m_d$ and $m_\mu = m_s$ are not.

\subsection{Proton decay}
There are two sources of proton decay in the minimal ${\rm SU}(5)$ -- interactions mediated by the vector gauge bosons $X_\mu$ and $\overline{X}_\mu$, and processes involving the color triplet scalar $T=(3,1)_{-1/3}$ from the $5_H$. 

The vector gauge boson interactions with quarks and leptons arise from the fermion kinetic terms in the Lagrangian,
\bean
\mathcal{L}_{\rm kin}  = i \, {\rm Tr}\!\left(\overline{5^c} \,\slashed{D}\, 5^c\right) +  i \, {\rm Tr}\!\left(\overline{10} \,\slashed{D}\, 10\right)
\, \supset  \, g \ \overline{l} \,\slashed{X} \,d^c + g \ \overline{q} \,\slashed{X} \,e^c + g \ \overline{u^c} \,\slashed{X} \, q \, + {\rm h.c.}.
\eean
Those terms give rise to  dimension-six operators mediating proton decay
\bean
\mathcal{L}^{(X)}_{\rm dim\,6} = \frac{g^2}{m_X^2} \left(\,\overline{u^c} \gamma_\mu q\,\right)\left(\,\overline{e^c}\gamma^\mu  q + \overline{d^c} \gamma^\mu l\,\right)\, + \,{\rm h.c.} \ ,
\eean
corresponding to the interaction shown in Fig.~1. 
\begin{figure}[t!]
\center
\includegraphics[width=0.45\linewidth]{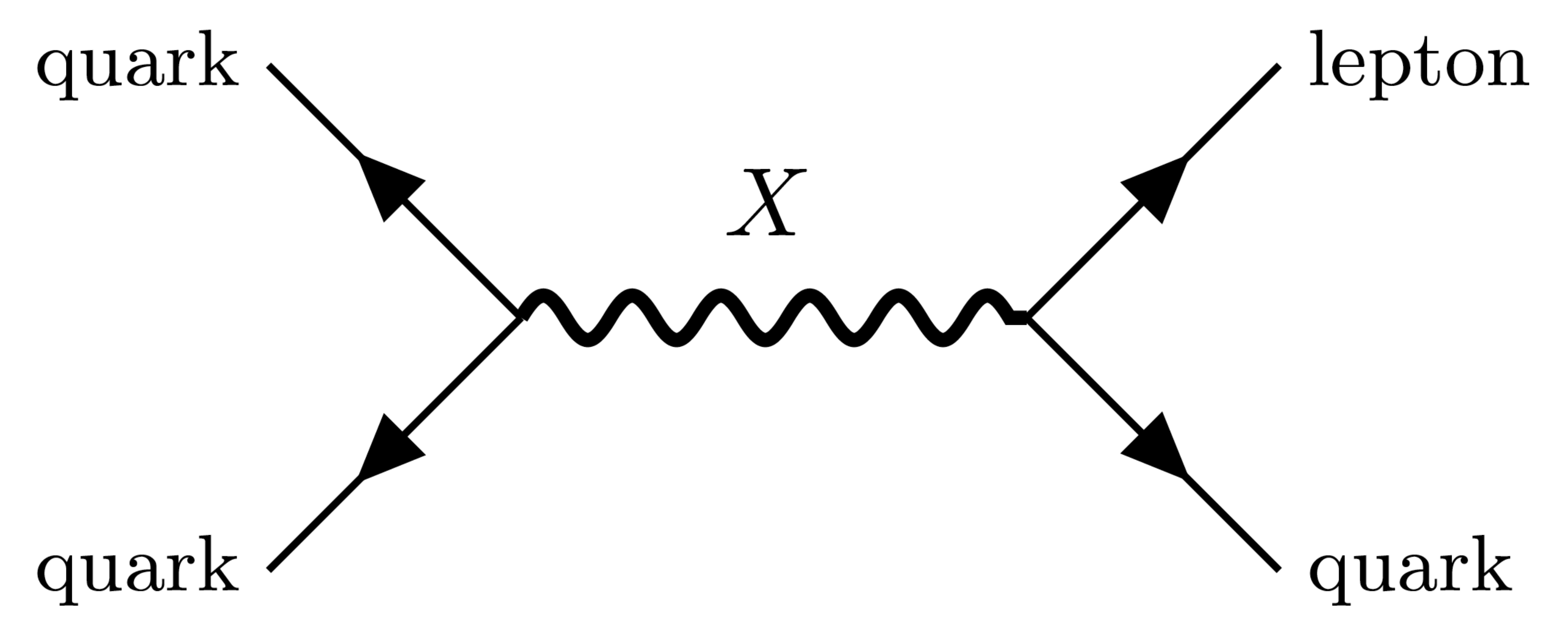} \vspace{-1mm}
\caption{\small{Proton decay mediated by the vector gauge boson $X_\mu = (3,2)_{-5/6}$\,.}}
\vspace{3mm}
\label{fig:1}
\end{figure}
The resulting proton decay rate is
$
\Gamma_p \sim {\alpha^2  m_p^5}/{m_{X}^4} 
$
and current  experimental limits on proton lifetime~\cite{Miura:2016krn} require

\vspace{-7mm}
\bean
m_X \gtrsim 10^{16} \ {\rm GeV}  \ .
\eean

The color triplet scalar interactions with quarks and leptons are described by the Yukawa terms 
\bean
\mathcal{L}_Y\, \supset   \, y_5 \ l\ T^*  q  +  y_5 \ d^c\, T^*  u^c + y_{10} \ q\, T \hspace{0.4mm}  q  + y_{10} \, u^c\, T\, e^c \, + \,{\rm h.c.}
\eean
and produce the dimension-six operators
\bean
\mathcal{L}^{(T)}_{\rm dim\,6} = \frac{y_5\,y_{10}}{m_T^2}\Big[\left(q\,q \right)\left(q\,l \right)+\left(d^c\,u^c \right)\left(u^c\,e^c \right)\Big]\, +\, {\rm h.c.}\,,
\eean
resulting in proton decay shown in Fig.~2. 
\begin{figure}[t!]
\center
\includegraphics[width=0.45\linewidth]{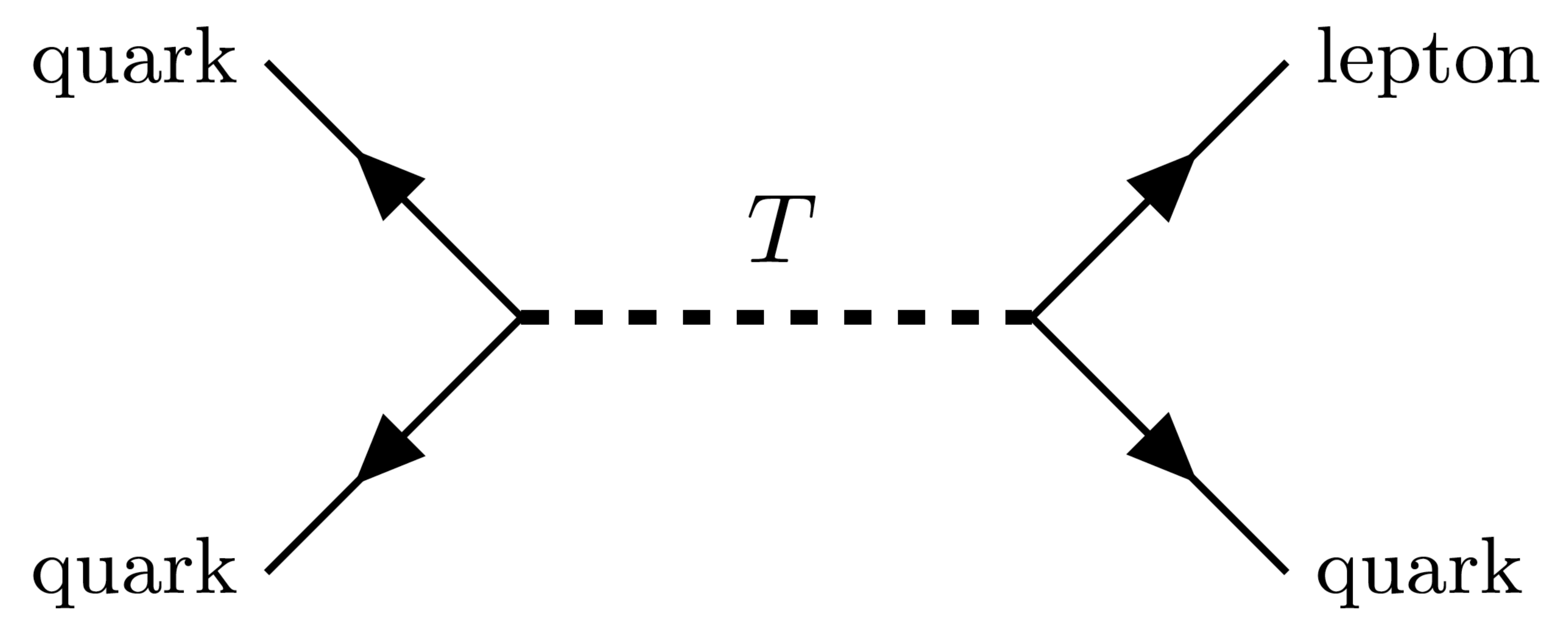} \vspace{0mm}
\caption{\small{Proton decay mediated by the scalar $T = (3,1)_{-1/3}$\,.}}
\vspace{3mm}
\label{fig:2}
\end{figure}
Because of the small Yukawa couplings, consistency with proton lifetime constraints leads to a less stringent bound on $m_T$ than the one on $m_X$, requiring merely
\bean
m_T \gtrsim 10^{12} \ {\rm GeV} \ .
\eean

We will show now how introducing extra fermion and scalar irreps into the minimal ${\rm SU}(5)$ GUT can forbid all proton decay channels discussed above, and how to forbid proton decay at any order in perturbation theory.

\section{SU(5) without proton decay}

We explicitly construct a four-dimensional non-supersymmetric ${\rm SU}(5)$ GUT in which the proton is stable. The idea is to add new irreps into the minimal ${\rm SU}(5)$ model and arrange that the physical SM quarks and leptons fall into different multiplets. The new ${\rm SU}(5)$ irreps introduced are $40$-plets and $50$-plets, since in their ${\rm SU}(3)_c\times {\rm SU}(2)_L \times {\rm U}(1)_Y$ decomposition they contain fields with the quantum numbers of SM quarks, but not the leptons. This allows to rotate the SM quark fields out of the $5$ and $10$ irreps, such that the leptons still reside in the $5$ and $10$, but the quarks themselves live entirely in the $40$'s and $50$'s. This arrangement prevents the vector gauge bosons $X_\mu$ and $\overline{X}_\mu$ as well as the scalar $T$ from connecting quarks to leptons.

\subsection{Fermion sector}
The new fermion irreps added to the minimal ${\rm SU}(5)$ model are two vector-like $40$-plets and two vector-like $50$-plets, so that the complete list of fermion irreps along with their ${\rm SU}(3)_c\times {\rm SU}(2)_L \times {\rm U}(1)_Y$ decomposition is \cite{Slansky:1981yr}:
\begin{eqnarray*}
 5^c \!\!\!\!\!\!\!&&= \, l \oplus D^c_{5} \ ,\nonumber\\ [3pt]
10 \!\!\!\!\!\!\! &&= \, e^c \oplus Q_{10}\oplus U^c_{10} \ ,\nonumber\\ [4pt]
40_{i} \!\!\!\!\!\!\!\!&&= \,Q_{40_i } \!\oplus U^c_{40_i } \!\oplus (1,2)_{-{3}/{2}}   \oplus (\bar{3}, 3)_{-{2}/{3}}  \oplus (8,1)_1 \oplus (\bar{6}, 2)_{{1}/{6}}  \ , \nonumber\\[4pt]
\overline{{40}}_{{i}} \!\!\!\!\!\!\!\!&&= \,Q_{\overline{40}_{i}}^c\! \oplus {{U_{\overline{40}_i}}} \!\oplus (1,2)_{{3}/{2}}   \oplus ({3}, 3)_{{2}/{3}}  \oplus (8,1)_{-1} \oplus ({6}, 2)_{-{1}/{6}}   \ ,\nonumber\\[4pt]
50_{i}^c \!\!\!\!\!\!\!\!&&= D_{50_i}^c  \oplus(1,1)_{2}   \oplus (3, 2)_{{7}/{6}} \oplus (6, 3)_{{1}/{3}}  \oplus (\bar{6},1)_{-{4}/{3}}  \oplus  (8, 2)_{-{1}/{2}}\ ,\nonumber\\[4pt]
\overline{{50^c_{i}}}\!\!\!\!\!\!\!\!&&= {D_{\overline{50}_i}}  \oplus(1,1)_{-2}   \oplus (\bar{3}, 2)_{-{7}/{6}} \oplus (\bar{6}, 3)_{-{1}/{3}}  \oplus (6,1)_{{4}/{3}}  \oplus  (8, 2)_{{1}/{2}} \ ,
\end{eqnarray*}
where $i=1,2$. Note that  $D^c_{5}$, $Q_{10}$ and $U^c_{10}$ are not the SM quark fields -- they mix with the fields in the same SM representation residing in other ${\rm SU}(5)$ multiplets, and the SM quarks are their linear combinations. The full decomposition including ${\rm SU}(3)_c$ and ${\rm SU}(2)_L$ indices is given in the appendix. 

\subsection{Higgs sector and symmetry breaking}
In the scalar sector, instead of the usual $5_H$ and $24_H$, one introduces the irreps $24_H$, $45_H$  and $75_H$.  Their decomposition into SM multiplets is:
\bean
&&24_H \,=\, (1,1)_0 \oplus (1,3)_0 \oplus (3,2)_{-{5}/{6}} \oplus (\bar{3},2)_{{5}/{6}} \oplus (8,1)_0 \ ,\nonumber\\ [4pt]
&&45_H \,=\, H \oplus (3,1)_{-{1}/{3}} \oplus (3,3)_{-{1}/{3}} \oplus (\bar{3},1)_{{4}/{3}} \oplus  (\bar{3},2)_{-{7}/{6}} \oplus  (\bar{6}, 1)_{-{1}/{3}} \nonumber\\
&&\hspace{14mm} \oplus \ (8,2)_{{1}/{2}}  \ ,\nonumber\\ [3pt]
&&75_H\,=\, (1,1)_0 \oplus (3,1)_{{5}/{3}} \oplus (\bar3,1)_{-{5}/{3}}\oplus (3,2)_{-5/6} \oplus (\bar3,2)_{5/6}   \oplus   (\bar6,2)_{-5/6} \nonumber\\
&&\hspace{14mm} \oplus \ (6,2)_{5/6}  \oplus  (8,1)_0 \oplus (8,3)_0 \ .
 \eean
The irreps $24_H$ and $75_H$ acquire GUT-scale vevs, $v_{24}$ and $v_{75}$, which break ${\rm SU}(5) \rightarrow {\rm SU}(3)_c\times {\rm SU}(2)_L \times {\rm U}(1)_Y$ and, as explained below, provide GUT-scale masses to all beyond-SM fermions. The SM Higgs in the $45_H$  develops  the standard electroweak vev, breaking the electroweak symmetry down to electromagnetism and resulting in SM quark and lepton masses.\\

 The scalar potential of the theory, under the assumption of invariance under $24_H \rightarrow - 24_H$ and $75_H \rightarrow - 75_H$, is given by
 \bean
\mathcal{L}_H\!&\! =\!&\! - \ \tfrac{1}{2}\mu_{24}^2{\rm Tr} (24_H^2)  \!+\!\tfrac{1}{4} a_1\!\!\left[{\rm Tr} (24_H^2) \right]^2 \!+\!\tfrac{1}{4} a_2{\rm Tr} (24_H^4) \nonumber\\[2pt]
&&\!-\ \tfrac{1}{2}\mu_{75}^2{\rm Tr} (75_H^2) \! +\!\tfrac{1}{4}\sum   b_k  {\rm Tr} (75_H^4)_k + \tfrac12 \sum g_k  {\rm Tr} (24_H^275_H^2)_k\nonumber\\[2pt]
&&\!+ \ M_{45}^2{\rm Tr} \big(|45_H|^2\big) + \!\sum h_k  {\rm Tr} \big(24_H^2|45_H|^2\big)_k + ...\ ,
\eean
where $k=1,2,3$ correspond to contractions with the two lowest representations in a given trace combining into a singlet, 2-component tensor and 4-component tensor, respectively. The explicit index contractions are shown in the appendix.

There exists a large region of parameter space for which all components of the $24_H$ and $75_H$ have GUT-scale masses, apart from one linear combination of the $(3,2)_{-5/6}$ fields (from the $24_H$ and $75_H$) and one combination of the $(\bar3,2)_{5/6}$ fields, both remaining massless, since those are the would-be Goldstone bosons of the broken ${\rm SU}(5)$ \cite{Langacker:1980js,Hubsch:1984pg,Cummins:1985vg}. All components of the $45_H$  are naturally at the GUT scale and a tuning of parameters in the scalar potential is needed to reproduce the SM Higgs mass. This tuning is equivalent to the doublet-triplet splitting problem in the minimal ${\rm SU}(5)$ and perhaps can be avoided by introducing further ${\rm SU}(5)$ multiplets \cite{Grinstein:1982um,Masiero:1982fe}. 
 
\subsection{Fermion mass terms}
The Yukawa and pure mass terms in our model are:
\bean
\mathcal{L}_Y \!\!& =&\!\!   Y_l \, 5^c 10 \,45^*_H  +    Y_{u}^{ij} 40_i\,40_j \,45_H + Y_d^{ij} 40_i\,50^c_j \,45^*_H +     M_{40}^{ij} \,\overline{40}_{i}\, {40_j}  \nonumber\\[1pt]
&+& \!\! \lambda^{ij}_{1}  24_H \overline{40}_{i} \,{40_j}+    \lambda^{ij}_{2}  \, \overline{40}_{i} \,24_H{40_j}  +  \lambda_{3}^{i} \, 24_H 10 \, \overline{40}_{i} + \lambda^{ij}_{4}  \, \overline{40}_{i} \,75_H{40_j} \nonumber\\[3pt]
 &+& \!\! \lambda_{5}^{i} \,75_H 10 \, \overline{40}_{i}   + M_{50}^{ij} \,{50^c_i}\, \overline{50^c_{j}}
+ \lambda^{ij}_{6} \,{50^c_i} \,24_H \overline{50^c_{j}} + \lambda^{ij}_{7} \,{50^c_i} \,75_H \overline{50^c_{j}} \nonumber\\[2pt]
&+&  \!\!  \lambda_{8}^{i}  \,75_H 5^c \, \overline{50^c_{i}} + {\rm h.c.}  \, ,
\eean
where $i, j = 1,2$ and the coefficients of the only other allowed contractions $10\ 40_i \,45_H$ are tuned to zero. We will now show that there exists a region of parameter space for which all new fermions have masses at the GUT scale, and at the same time all masses of the SM particles can be recovered. \\

Focusing on the fields with the quantum numbers of the SM down quark, after ${\rm SU}(5)$ breaking the relevant mass terms are
\bean
\mathcal{L}_{\rm mass} = \left(\begin{matrix} \, {D_{\overline{50}_1}} & {D_{\overline{50}_2}} \, \end{matrix}\right)
\mathcal{M}_D \!\left( \begin{matrix}
D^c_{5} \ \\
\,D^c_{50_1}\\
D^c_{50_2} 
\end{matrix} \right) \, .
\eean
Performing a biunitary transformation to the mass eigenstate basis, 
$\mathcal{M}^{\rm diag}_D = ({R}_D)_{2\times2} \,\mathcal{M}_{D} \, ({L}_{D})^\dagger_{3\times3} $, the mass eigenstates are
\bean
\left( \begin{matrix}
\,{D^c_1}\,\\
{D^c_{2}} \\
{D^c_{3}}
\end{matrix} \right) ={L}_D\!\left( \begin{matrix}
D^c_{5} \ \\
\,D^c_{50_1}\\
D^c_{50_2} 
\end{matrix} \right).
\eean
In order to rotate the SM down quark out of the $5^c$ irrep, it is sufficient for the mass eigenstate ${D^c_1}$ not to contain any admixture of ${D^c_{5}}$. This is accomplished by imposing the condition
\bean
{\rm det}\left({M_{50}^{ij}}+\tfrac{1}{3\sqrt{30}}\lambda_{6}^{ij}v_{24}  +\tfrac{1}{3\sqrt2}\lambda_{7}^{ij}v_{75}\right)  = 0 \ .
\eean
This tuning of parameters guarantees that the SM down quark field $d^c$ resides only in the $50^c_1$ and $50^c_2$ irreps, i.e.,
\bean
d^c \equiv D^c_1 = {L}_D^{12} D^c_{50_1} +{L}_D^{13} D^c_{50_2} \ ,
\eean
where the coefficients ${L}_D^{12}$ and ${L}_D^{13}$ are functions of the  Lagrangian parameters.
This ensures that $d^c$ does not get its mass from ${\rm SU}(5)$ breaking. 
An explicit calculation reveals that for the above choice of parameters all other fields in the $50^c_1$ and $50^c_2$ have GUT-scale masses.

The same strategy can be applied to the SM quark doublet and the up quark.  The physical $q$ and $u^c$ are rotated out of the $10$ irrep and  end up as linear combinations of the corresponding fields from the $40_1$ and $40_2$ irreps. Again, it can be shown that all other fields in the $40_1$ and $40_2$ develop masses at the GUT scale.

Ultimately, the SM quark and lepton masses originate entirely from electroweak symmetry breaking through the Lagrangian terms
\bean
{\mathcal{L}}_Y \!\!&\supset&\!\! \,Y_l \,5^c 10 \,45^*_H + Y_{u}^{ij} \,40_i\,40_j \,45_H + Y_d^{ij}\,40_i\,50^c_j \,45^*_H  + {\rm h.c.}\nonumber\\
\hspace{10mm}\!\!&\supset& \!\! \,y_l \,l\, H^* e^c + y_u \,q\, H\, u^c  + y_d \,q \, H^* d^c  + {\rm h.c.}\ .
\eean
Contrary to the minimal ${\rm SU}(5)$ scenario, there is no problematic relation between the electron and down quark masses.

\subsection{Proton stability}

\subsubsection{Tree level}
\vspace{-1mm}
The most dangerous proton decay operators in the standard ${\rm SU}(5)$ GUT arise from fermion kinetic terms, as discussed earlier, and involve the vector gauge bosons $X_\mu = (3,2)_{-5/6}$ and $\overline{X}_\mu = (\bar3,2)_{5/6}$. In our model, the corresponding Lagrangian terms are 

\vspace{-7mm}
\bean
\mathcal{L}_{\rm kin}  =\, i \sum_R {\rm Tr}\left(\overline{R} \,\slashed{D}\, R\right)  ,
\eean

\vspace{-1mm}
\noindent
with  $R=$ $5^c$, $10$, $40_i$, $\overline{40}_{i}$, $50^c_i$ and $\overline{50^c_{i}}$. However, since the SM leptons live in the $5$ and $10$ irreps, whereas the SM quarks live in the $40$ and $50$ irreps, in our model there are no vertices connecting $X_\mu$ or $\overline{X}_\mu$ to a quark and a lepton. This immediately implies that there is no tree-level proton decay through a vector gauge boson exchange.

It is also straightforward to check that our model is free from tree-level proton decay mediated by scalars. For the same reasons as above, the terms 

\vspace{-6mm}
\bean
{\mathcal{L}}_Y \supset \,Y_l \,5^c 10 \,45^*_H + Y_{u}^{ij} \,40_i\,40_j \,45_H + Y_d^{ij}\,40_i\,50^c_j \,45^*_H  + {\rm h.c.}
\eean

\vspace{-1mm}
\noindent
do not result in any vertices connecting the color triplet scalar $T=(3,1)_{-1/3}$ or any other scalar from the $45$ irrep to a quark and a lepton. This completes the proof that there is no tree-level proton decay in our model.

\subsubsection{Loop level}
\vspace{-1mm}
To investigate proton decay at higher orders in perturbation theory, it is no longer possible to do this on a case by case basis, and a symmetry argument is needed. It turns out that  our model does exhibit such a partial discrete symmetry -- all Lagrangian terms, apart from $\lambda_3^i24_H 10 \, \overline{40}_{i}$, $\lambda_5^i75_H 10 \, \overline{40}_{i}$ and  $\lambda_8^i75_H 5^c \, \overline{50^c_{i}}$, are invariant upon substituting
\bean
5^c \rightarrow - 5^c \ ,& \ \  10 \rightarrow -10 \ . \ 
\eean
Under this transformation the SM leptons are odd, since they live in the $5$ and $10$ irreps, whereas the quarks are even, since they reside in other irreps. In proton decay the initial state involves no leptons and no heavy states, so it is even under this transformation, whereas the final state consists of an odd number of leptons and no heavy states, so it is odd. This implies that proton decay is forbidden at any loop order as long as the fields from the $24_H$ and $75_H$ are not involved.  
One cannot set $\lambda_3^i \!=\! \lambda_5^i \!=\! \lambda_8^i \!=\! 0$ to remove the terms not invariant under $5^c \rightarrow - 5^c , \  10 \rightarrow -10$, since then it would be impossible to rotate the SM quarks out of the $5$ and $10$ irreps.
To forbid the remaining proton decay channels we assume that ${\rm SU}(5)$ breaking is non-linearly realized \cite{Coleman:1969sm}. The components of $24_H$ and $75_H$ decouple and at the Lagrangian level they are replaced by non-dynamical condensates. The scalar sector of the theory is then described by a nonlinear sigma model \cite{GellMann:1960np,Callan:1969sn}.

Let us note that an alternative recent proposal \cite{Karananas:2017mxm} uses the same argument to remove the $24_H$ fields from the spectrum of the minimal ${\rm SU}(5)$  GUT. That model, however, achieves proton stability by imposing specific gauge conditions that eliminate all beyond-SM fields from the theory, making it indistinguishable from the SM. The only other attempts to construct 4D GUT models based on a single gauge group without proton decay we are aware of \cite{Fritzsch:1975wn,GellMann:1976pg,Langacker:1977ai,Langacker:1978fn,Segre:1980qc,Kuzmin:1981ek,Fayet:1984fe,Berezhiani:2001ub} are either  experimentally excluded by now due to
the presence of new light particles with SM charges or suffer from tree-level proton decay mediated by scalars that cannot be removed by invoking non-linear symmetry breaking.
\vspace{-2mm}

\section{Conclusions} 
\vspace{-1mm}
We have constructed a four-dimensional grand unified theory based on ${\rm SU}(5)$ that does not suffer from proton decay at any order in perturbation theory. The idea is to separate the physical quark and lepton fields into different representations of the gauge group. The absence of proton decay at tree level is achieved by adding extra multiplets into the theory and imposing specific relations between the model parameters. Full proton stability requires nonlinear ${\rm SU}(5)$ breaking.

Another interesting feature of the model is the possibility of having full gauge coupling unification, despite the theory being non-supersymmetric. This can be realized by lowering the masses of some of the scalars in the $45_H$ to the TeV scale and adding one more scalar representation \cite{Stone:2011dn,Murayama:1991ah,Cox:2016epl}. This provides the opportunity to test the model at the LHC.

Although our specific construction is based on ${\rm SU}(5)$, it is meant to serve only as a proof of concept that grand unified theories built on a single gauge group with a stable proton do exist. Perhaps a simpler and more attractive theory of this type can be constructed in the framework of the gauge group ${\rm SO}(10)$. We hope that our finding will revive the interest in grand unification and open the door to a new branch of model building.

\subsection*{Acknowledgments}
B.F. would like to thank the organizers of the Conference on Particles and Cosmology in Singapore, especially the
chairman, Harald Fritzsch, for the invitation, warm hospitality, and a wonderful
scientific atmosphere.
This research was supported in part by the DOE Grant No.~${\rm DE}$-${\rm SC0009919}$. \vspace{5mm}

\appendix
\section{SU(5) representations}
Below we provide the ${\rm SU}(3)\times {\rm SU}(2) \times {\rm U}(1)$ decomposition of the ${\rm SU}(5)$ multiplets relevant for our model. The $\alpha, \beta, \gamma, \delta, \sigma$ are ${\rm SU}(3)$ indices  and $a, b, c, d$ are ${\rm SU}(2)$ indices:\vspace{4mm}
\bean
&&5^c_\alpha = ({D}^c_{5})_\alpha  \ , \ \ \ \ \ 5^c_a = \epsilon_{ab}{l}^b \ , \nonumber\\ [15pt]
&&10^{\alpha\beta}= \tfrac{1}{\sqrt2}\epsilon^{\alpha\beta\gamma} (U_{10}^c)_\gamma \ , \ \ \ \ 10^{\alpha a} = -\tfrac{1}{\sqrt2} Q^{\alpha a}_{10} \ , \ \ \ \ 
 10^{ab} = \tfrac{1}{\sqrt2}\epsilon^{ab} e^c \ ,  \nonumber\\ [15pt]
  &&24^\alpha_\beta =  [(8,1)_0]^{\alpha}_\beta + \tfrac{2}{\sqrt{30}}\delta^\alpha_\beta (1,1)_0   \ , \  \ \  \ 24^\alpha_a = \tfrac{1}{\sqrt{2}}  [(3,2)_{-\frac56}]^{\alpha}_a   \ ,\nonumber\\
 &&24^a_b =  [(1,3)_0]^{a}_b - \tfrac{3}{\sqrt{30}}\delta^a_b (1,1)_0   \ , \ \ \  \ 
  \langle \,24^A_B \,\rangle = \tfrac{1}{\sqrt{30}}v_{24} \left(2\delta^\alpha_{\beta}-3\delta^a_{b}\right)  , \nonumber\\ [15pt]
&&40^{\alpha\beta\gamma}_\delta  =\tfrac13 \epsilon^{\alpha\beta\gamma} (U^c_{40})_\delta \ , \ \ \ \ \ 40^{\alpha\beta \gamma}_a = \tfrac{1}{\sqrt6}\epsilon^{\alpha\beta\gamma} [(1,2)_{-\frac32}]_a  \ , \nonumber\\
&&40^{\alpha\beta a}_\gamma =   -\tfrac13\delta^{[\alpha}_{\gamma}(Q_{40})^{\beta] a} + \tfrac{1}{\sqrt6}\epsilon^{\alpha\beta\delta}[(\bar6,2)_{\frac16}]_{\gamma\delta}^a \ , \nonumber\\
&&40^{\alpha\beta a}_b = -\tfrac16 \epsilon^{\alpha\beta\gamma}\delta^a_b (U^c_{40})_\gamma   +\tfrac{1}{\sqrt6} \epsilon^{\alpha\beta\gamma}[(\bar3,3)_{-\frac23}]^a_{b\gamma} \ ,\nonumber\\
&&40^{\alpha a b}_\beta =  \tfrac{1}{\sqrt6}\epsilon^{ab}[(8,1)_{1}]^{\alpha}_{\beta} \ , \ \ \ \ \
40^{\alpha a b}_c= \tfrac13\epsilon^{ab} (Q_{40})^\alpha_c \ ,\nonumber\\ [15pt]
&&45^{\alpha\beta}_{\gamma} = \tfrac{1}{\sqrt2}\epsilon^{\alpha\beta\delta}[(\bar6,1)_{-\frac13}]_{\delta\gamma}+  \tfrac{1}{\sqrt2} \delta^{[\alpha}_\gamma [(3,1)_{-\frac13}]^{\beta]} \ ,\nonumber\\
&&45^{\alpha\beta}_{a} = \tfrac{1}{\sqrt2}\epsilon^{\alpha\beta\gamma}[(\bar3,2)_{-\frac76}]_{\gamma a} \ , \ \ \ \ \
45^{\alpha a}_{\beta} =  \tfrac{1}{\sqrt2}[(8,2)_\frac12]^{\alpha a}_\beta + \tfrac{1}{2\sqrt6}\delta^\alpha_\beta H^a \ , \nonumber\\
&&45^{a\alpha}_{b} =  \tfrac{1}{\sqrt2}[(3,3)_{-\frac13}]^{a \alpha}_b  -\tfrac{1}{2\sqrt2}\delta^a_b[(3,1)_{-\frac13}]^\alpha \ ,\nonumber\\
&&45^{ab}_{\alpha} =  \tfrac{1}{\sqrt2}\epsilon^{ab}[(\bar3,1)_{\frac43}]_{\alpha} \ , \ \ \ \ \ 
45^{ab}_{c} = - \tfrac{3}{\sqrt6} \delta^{[a}_cH^{b]} \ ,
\eean
\bean
&&\!\!\!\!\!\!\!\!50^{\alpha\beta\gamma}_{\delta\sigma} = \tfrac{1}{3}\delta^{[\alpha}_\delta \delta^{\beta}_\sigma D_{50}^{\gamma]} \, , \ \
 50^{\alpha\beta a}_{\gamma\delta} = \tfrac{2}{\sqrt6} \delta^{[\alpha}_{[\gamma}[(8,2)_\frac{1}{2}]^{\beta] a}_{\delta]} , \ \
 50^{\alpha\beta \gamma}_{a b} = \tfrac{1}{2\sqrt3}\epsilon^{\alpha\beta\gamma}\epsilon_{ab} (1,1)_{-2} \, ,\nonumber\\
&&\!\!\!\!\!\!\!\!50^{\alpha\beta \gamma}_{\delta a} =\tfrac{1}{2\sqrt6}\Big[ \delta^{[\gamma}_\delta \epsilon^{\alpha\beta]\sigma}[(\bar{3},2)_{-\frac76}]_{\sigma a} \, , \ \ 
 50^{\alpha \beta a}_{\gamma b} =\tfrac{1}{2\sqrt3} \epsilon^{\alpha\beta\delta}[(\bar6,3)_{-\frac13}]^{a}_{\gamma \delta b} + \tfrac{1}{6} \delta^a_b \delta^{[\alpha}_\gamma D_{50}^{\beta]}\, , \nonumber\\
&&\!\!\!\!\!\!\!\!50^{ \alpha\beta a}_{b c} =-\tfrac{1}{\sqrt6} \epsilon^{\alpha\beta\sigma}\delta^a_{[b}[(\bar3,2)_{-\frac76}]_{c]\sigma}\ , \ \ \,
 50^{\alpha a b}_{\beta\gamma} = \tfrac{1}{2\sqrt3} \epsilon^{ab}\epsilon_{\beta\gamma\delta}[(6,1)_{\frac43}]^{\alpha\delta}\, ,\nonumber\\
 &&\!\!\!\!\!\!\!\!50^{ a b \alpha}_{c \beta} =\tfrac{1}{\sqrt6}\delta^{[a}_c [(8,2)_{\frac12}]^{b]\alpha}_\beta  \ , \ \ \ \ 
  50^{ a b \alpha}_{c d} = \tfrac13 \delta^{[a}_{c} \delta^{b]}_{d} D_{50}^\alpha \ ,\nonumber\\[15pt]
 &&\!\!\!\!\!\!\!\!75^{\alpha\beta}_{\gamma\delta} = -\tfrac{1}{\sqrt3}\delta^{[\alpha}_{\gamma} [(8,1)_0]^{\beta]}_{\delta}+\tfrac{1}{3\sqrt2} \delta^{[\alpha}_{\gamma} \delta^{\beta]}_{\delta}(1,1)_0 \ ,\nonumber\\
&&\!\!\!\!\!\!\!\!75^{\alpha\beta}_{\gamma a} =\tfrac{1}{2\sqrt2}[(\bar6,2)_{-\frac{5}{6}}]^{\alpha\beta}_{\gamma a} - \tfrac{1}{\sqrt6}\delta^{[\alpha}_{\gamma} [(3,2)_{-\frac56}]^{\beta]}_a \ , \ \ \ 75^{\alpha\beta}_{ab} = \tfrac12\epsilon^{\alpha\beta\gamma}\epsilon_{ab}[(\bar{3},1)_{-\frac53}]_\gamma \, , \nonumber\\
&&\!\!\!\!\!\!\!\!75^{\alpha a}_{\beta b} = \tfrac14 [(8,3)_0]^{\alpha a}_{\beta b} + \tfrac{1}{2\sqrt3} \delta^a_b[(8,1)_0]^{\alpha}_{\beta}  -  \tfrac{1}{6\sqrt2}\delta^\alpha_\beta\delta^a_b(1,1)_0 \ , \nonumber\\
 &&\!\!\!\!\!\!\!\!75^{a\alpha}_{bc} = \tfrac{2}{\sqrt6} \delta^a_{[b} [(3,2)_{-\frac56}]^\alpha_{c]} \ , \ \ \ \ \ 
75^{ab}_{cd} = \tfrac{1}{\sqrt2}\delta^{[a}_{c} \delta^{b]}_{d}(1,1)_0  \ ,\nonumber\\
&&\!\!\!\!\!\!\!\!\langle\,75^{AB}_{CD}\, \rangle = \tfrac{1}{3\sqrt2} v_{75}  \big(\delta^{[\alpha}_{\gamma}\delta^{\beta]}_{\delta} + 3\delta^{[a}_{c}\delta^{b]}_{d} -  \tfrac{1}{2} \delta^{\alpha}_{\gamma}\delta^{b}_{d} + \tfrac{1}{2} \delta^{\alpha}_{\delta}\delta^{b}_{c}  +  \tfrac{1}{2} \delta^{\beta}_{\gamma}\delta^{a}_{d} -  \tfrac{1}{2} \delta^{\beta}_{\delta}\delta^{a}_{c}  \big) \ .
\eean

\vspace{-2mm}

\section{Scalar potential}
Upon writing the indices out explicitly, the scalar potential takes the following form (subscript $H$ was dropped for clarity),
\bean
\mathcal{L}_H\!\!& =&\!\! - \ \tfrac{1}{2}\mu_{24}^2 24^i_j 24^j_i  + \tfrac{1}{4} a_1(24^i_j 24^j_i )^2 \!+\!\tfrac{1}{4} a_2 24^i_j 24^j_k 24^k_l 24^l_i   -  \tfrac{1}{2}\mu_{75}^2 75^{ij}_{kl} 75^{kl}_{ij} \nonumber\\
&&\!\!+ \ \tfrac{1}{4}  b_1 (75^{ij}_{kl} 75^{kl}_{ij})^2 +  \tfrac{1}{4}  b_2 75^{ij}_{pq} 75^{kl}_{ij}75^{mn}_{kl}75^{pq}_{mn}+\tfrac{1}{4}  b_3 75^{ij}_{kl} 75^{kl}_{in}75^{mn}_{pq}75^{pq}_{mk}\nonumber\\
&&\!\! + \  \tfrac{1}{4} c_1 (24^i_j 24^j_i)(75^{kl}_{mn} 75^{mn}_{kl})  +\tfrac14 c_2 \, 24^i_j 24^j_k 75^{kl}_{mn} 75^{mn}_{li} \nonumber\\
&&\!\!+ \   \tfrac{1}{4} c_3 \,24^i_j 75^{jk}_{il} 24^m_n 75^{nl}_{mk}   +   M_{45}^2\,\overline{45}^i_{jk} 45^{jk}_i  +    d_1 (24^i_j 24^j_i)(\overline{45}^{k}_{lm} 45^{lm}_{k}) \nonumber\\
&&\!\!+ \ d_2 \, 24^i_j 24^j_k\,\overline{45}^{k}_{lm} 45^{lm}_{i}   +  d_3\,24^i_j \,\overline{45}^{j}_{ik} 24^l_m 45^{mk}_{l}  + \sum e_i \, {\rm Tr} (75^245^2)_i \ .
\eean

\bibliographystyle{utphys}
\bibliography{SU5}

\end{document}